# The nonextensive parameter for nonequilibrium plasmas in magnetic field

Yu Haining and Du Jiulin

*Department of Physics, School of Science, Tianjin University,Tianjin 300072, China*

**Abstract** – The nonextensive parameter for nonequilibrium electron gas in a magnetic field of plasma is studied. We exactly obtained an expression of the *q*-parameter based on Boltzmann kinetic theories for plasmas, where Coulombian interactions and Lorentz forces play dominant roles. We show that the *q*-parameter different from unity is closely related to the temperature gradient, the electric field strength, the magnetic induction as well as the overall bulk velocity of the gas. The effect of the magnetic field on the *q*-parameter depends on the overall bulk velocity of the gas. Thus the nonextensive *q*-parameter in magnetic field represents the nonequilibrium nature or nonisothermal configurations of nonequilibrium electron gas in the electromagnetic interactions.

Nonextensive statistical mechanics (NSM) [1] has been attracting more and more attention since it has been successfully applied to study a variety of interesting problems in natural science. In physics, self-gravitating systems and plasma systems have offered the best framework for searching into the nonextensive effects as well as the non-Maxwellian distributions because the long-range interactions between particles play a fundamental role in governing the dynamical properties of such systems. In NSM, the velocity distribution is described by a power-law *q*-distribution function, i.e. the generalized Maxwellian distribution [2],

$$f_q(\mathbf{v}) = nB_q \left(\frac{m}{2\pi kT}\right)^{3/2} \left[1-(1-q)\frac{m\mathbf{v}^2}{2kT}\right]^{1/(1-q)}, \quad (1)$$

where $B_q$ is a *q*-dependent normalization constant, *k* is Boltzmann's constant, *T* is temperature, *n* is particle number density, *m* is mass and **v** is velocity. *q* is the nonextensive parameter, $q \neq 1$ describes nonextensivity of the system under consideration, and in the limiting $q \to 1$ eq.(1) recovers a Maxwellian velocity distribution. Recently, this power-law *q*-distribution could be exactly generated from the stochastic dynamics of the Brownian motion by introducing a new fluctuation-dissipation relation [3-5].

It is undoubtedly very important to correctly understand the physical meaning of the nonextensive *q*-parameter different from unity in the power-law *q*-distribution. For self-gravitating gas, one has known that the *q*-parameter is related to gravitational potential $\varphi$ and temperature gradient $\nabla T$ in the gas by the equation [6], $k\nabla T + (1-q)m\nabla\varphi = 0$, which gives one physical explanation for the *q*-parameter different from unity, and tells us a fact that the power-law *q*-distribution function represents a stationary nonequilibrium distribution. When being applied to the solar interior, this equation received the experimental support from helioseismological measurements [7], and found the convective stability criterion [8]. This equation can also be applied to determine thermodynamic stability criterion [9] and convective stability criterion for a general self-gravitating gas [10]. In addition, for the nonextensive systems with self-gravitating long-range interactions, one finds that the *q*-parameter is associated with the gravitational potential $\varphi$ and the velocity dispersion $\sigma$ by the equation [11], $2\sigma\nabla\sigma + (1-q)\nabla\varphi = 0$, and thus it represents characteristics of the long-range interactions and the non-local correlations within the systems. Using this equation and the corresponding *q*-distribution, one has excellently modeled the dark matter haloes observed in spherical galaxies with regard to their nonequilibrium stationary states [12]. Most recently, the *q*-parameter in an actual gas is studied using the second virial coefficient and thus associated with the interaction potential between molecules [13].

Similar to a self-gravitating gas, the *q*-parameter for the plasma without magnetic field is related to Coulombian interaction potential $\varphi$ and temperature gradient $\nabla T$ in the plasma by the equation [14], $k\nabla T - e(1-q)\nabla\varphi = 0$, which thus show us one clear physical interpretation for the *q*-parameter different from unity in the plasmas. One has understood that the power-law *q*-distribution in plasmas does not represent a thermal equilibrium state but a stationary nonequilibrium state. NSM has been applied to many investigations of physical properties of the astrophysical and space plasmas with power-law distributions, such as solar wind and suprathermal tails [15-18], waves and instabilities [19-43], charging phenomena [44-49] and some other properties [50-54] *etc*. However, the equation of *q*-parameter above is only for the plasmas without magnetic field. One has not known yet the equation of *q*-parameter for the plasmas in magnetic field. Generally, there are electromagnetic interactions between plasmas and magnetic fields, which, as pointed out by Alfven [55], are important for space physics and astrophysics. Based on this theory, magneto-hydrodynamics has been widely applied to study plasma physics of magnetosphere, interplanetary space and solar physics *etc*.

Early in 1968, based on using observations of low-energy electrons in the magnetosphere, Vasyliunas analyzed energy spectra of observed electrons within the plasma sheet and drew an empirical function to model the velocity distribution of the electrons [56]. This is a power-law distribution with a kappa parameter and therefore it is later known as the *κ*-distribution. In fact, spacecraft measurements of plasma velocity distributions, both in the solar wind and in planetary magnetosheaths, have revealed that non-Maxwellian distributions are quite common. In many situations the distributions appear reasonably Maxwellian at low energies but have a "suprathermal" power-law tail with the *κ*-distribution at high energies. In the solar corona, *κ*-like power-law distributions have been proposed to arise from strong nonequilibrium thermodynamic gradients, and temperature anisotropies in the solar wind are correlated with the magnetic fields (see [57] and the references therein). In such circumstances, if one want to study properties of the plasmas with power-law *q*-distribution, one need to consider the role of magnetic field. In this letter, we study the equation of *q*-parameter for the plasma in magnetic field.

**Basic equation.** – Following the idea of ref.[14], generally we consider a class of spatially inhomogeneous plasma gas with the overall bulk velocity **u** of the gas. The power-law *q*-distribution eq.(1) can be rewritten as

$$f_q(\boldsymbol{r},\boldsymbol{v}) = n(\boldsymbol{r})B_q\left(\frac{m}{2\pi kT(\boldsymbol{r})}\right)^{3/2}\left[1-(1-q)\frac{m(\boldsymbol{v}-\boldsymbol{u})^2}{2kT(\boldsymbol{r})}\right]^{1/(1-q)}. \quad (2)$$

As usual, there is a thermal cutoff on the maximum value allowed for the velocity of a particle (an electron or ion) for $q < 1$, $|\boldsymbol{v}-\boldsymbol{u}|_{\max} = \sqrt{2kT/m(1-q)}$, whereas for $q > 1$ without the restriction. In fact, the power-law distribution eq.(2) is equivalent to the *κ*-distribution known in the space plasma and astrophysics if we make a parameter transition simply using $q-1=\kappa^{-1}$, and the character of *q* (or *κ*) for $q > 1$ and $q < 1$ is manifest in the dual nature of interactions, showing important roles of the nonextensive parameter in the understanding of the gravitational and electromagnetic interactions for dark matter and interplanetary plasmas [15,16, 58].

In the plasma kinetic theory, many plasma processes are analyzed by using the Boltzmann equation, such as transport processes etc [54, 59, 60]. We consider a general plasma gas, where electrons are moving in an electromagnetic field, and thus they are driven by both the electric field force and the Lorentz force. Without loss generality, we use $f_q(\boldsymbol{r},\boldsymbol{v},t)$ to denote the time-dependent velocity distribution function of electrons, while the distribution of ions is considered constant. Then, the kinetic behaviour is governed by the generalized Boltzmann equation [61],

$$\frac{\partial f_q}{\partial t} + \boldsymbol{v}\cdot\frac{\partial f_q}{\partial \boldsymbol{r}} + \frac{e}{m}\left(\nabla\varphi - \frac{\boldsymbol{v}}{c}\times\boldsymbol{B}\right)\cdot\frac{\partial f_q}{\partial \boldsymbol{v}} = C_q(f_q), \quad (3)$$

where $C_q$ is the *q*-collision term, *e* is charge of an electron, *c* is the light speed, **B** is the magnetic induction (does not depend explicitly on time), and $\varphi$ is the electric potential and satisfies the Poisson equation, $\nabla^2\varphi = 4\pi en$.

The so-called generalized Boltzmann equation is the Boltzmann equation that has the equilibrium solution of the power-law *q*-distribution (2). It has been proved [61] that the equilibrium solutions of eq. (3) satisfy the *q*–H theorem only if $q > 0$ and evolve irreversibly towards the power-law *q*-distribution (2) (*i.e.* the generalized Maxwellian *q*-distribution [2]). Namely, on the basis of the *q*–H theorem, the distribution $f_q(\boldsymbol{r},\boldsymbol{v},t)$ will evolve irreversibly towards the power-law *q*-distribution (2) so that the plasma will reach a stationary equilibrium state. At this time, the collision term $C_q$ vanishes and eq.(3) naturally becomes

$$\boldsymbol{v}\cdot\nabla f_q + \frac{e}{m}\left(\nabla\varphi - \frac{\boldsymbol{v}}{c}\times\boldsymbol{B}\right)\cdot\nabla_v f_q = 0, \quad (4)$$

where we have used $\nabla = \partial/\partial\boldsymbol{r}$ and $\nabla_v = \partial/\partial\boldsymbol{v}$.

The conclusion that the power-law *q*-distribution (2) is an equilibrium solution of the Boltzmann equation is not only a theoretical result, but also an experimental result observed in space plasmas. In fact, the power-law *q*-distribution is equivalent to the *κ*-distribution and the *κ*-like power-law distributions noted in the spacecraft measurements of plasma velocity distributions both in the solar wind and in planetary magnetospheres and magnetosheaths (see [3, 5], [54], [57] and the references therein), and therefore it has been so widely applied to study properties of the astrophysical and space plasmas with power-law distributions [14-53]. Now the kappa-distribution family has been frequently discussed and studied by using the $q$-distributions under the framework of nonextensive statistics.

**The property of the $q$-parameter.** – In order to study the property of the $q$-parameter (or $\kappa$-parameter) as well as the power-law $q$-distribution (2), we can write eq.(4) in another form as

$$\mathbf{v} \cdot \nabla \ln f_q + \frac{e}{m}\left(\nabla \varphi - \frac{\mathbf{v}}{c} \times \mathbf{B}\right) \cdot \nabla_v \ln f_q = 0 \cdot \quad (5)$$

The logarithmic form of eq. (2) can be shown as the power series by

$$\ln f_q = \ln\left[nB_q\left(\frac{m}{2\pi kT}\right)^{3/2}\right] - \sum_{i=1}^{\infty}\frac{1}{i}(1-q)^{i-1}\left[\frac{m(\mathbf{v}-\mathbf{u})^2}{2kT}\right]^i, \quad (6)$$

which is defined only under the condition of $-1 \le (1-q)m(\mathbf{v}-\mathbf{u})^2/2kT < 1$. This condition is equivalent to a thermal cutoff on the maximum value allowed for the velocity of an electron for $q < 1$, i.e. $|\mathbf{v}-\mathbf{u}| < \sqrt{2kT/m(1-q)}$, and for $q > 1$, $|\mathbf{v}-\mathbf{u}| \le \sqrt{2kT/m(q-1)}$. Substituting eq.(6) into eq.(5), we have

$$\mathbf{v} \cdot \left\{\nabla \ln\left[nB_q\left(\frac{m}{2\pi kT}\right)^{3/2}\right] + \sum_{i=1}^{\infty} \mathbf{A}_{qi}\left[\frac{m(\mathbf{v}-\mathbf{u})^2}{2kT}\right]^i\right\}$$
$$-\frac{e\nabla \varphi}{kT} \cdot (\mathbf{v}-\mathbf{u}) \sum_{i=0}^{\infty}\left[\frac{(1-q)m}{2kT}(\mathbf{v}-\mathbf{u})^2\right]^i$$
$$+\frac{e}{kT}\frac{1}{c}(\mathbf{v}\times\mathbf{B})\cdot(\mathbf{v}-\mathbf{u})\sum_{i=0}^{\infty}\left[\frac{(1-q)m}{2kT}(\mathbf{v}-\mathbf{u})^2\right]^i = 0, \quad (7)$$

where $\mathbf{A}_{qi}$ denotes

$$\mathbf{A}_{qi} = (1-q)^{i-1}\frac{1}{T}\nabla T \cdot \quad (8)$$

Expanding $(\mathbf{v}-\mathbf{u})^{2i}$ in eq.(7), and using the formula, $(\mathbf{a}\times\mathbf{b})\cdot\mathbf{c} = (\mathbf{c}\times\mathbf{a})\cdot\mathbf{b} = (\mathbf{b}\times\mathbf{c})\cdot\mathbf{a}$, we have

$$\mathbf{v} \cdot \nabla \ln\left[nB_q\left(\frac{m}{2\pi kT}\right)^{3/2}\right] + \mathbf{v} \cdot \sum_{i=1}^{\infty} \mathbf{A}_{qi}\left(\frac{m}{2kT}\right)^i$$
$$\sum_{p+s+j=i}\frac{i!}{p!s!j!}(-2\mathbf{v}\cdot\mathbf{u})^s u^{2j}v^{2p}$$
$$-\left[\frac{e\nabla\varphi}{kT}\cdot(\mathbf{v}-\mathbf{u}) - \frac{e}{kT}\frac{1}{c}(\mathbf{u}\times\mathbf{B})\cdot\mathbf{v}\right]$$
$$\sum_{i=0}^{\infty}\left[\frac{(1-q)m}{2kT}\right]^i \sum_{p+s+j=i}\frac{i!}{p!s!j!}(-2\mathbf{v}\cdot\mathbf{u})^s u^{2j}v^{2p} = 0. \quad (9)$$

Because $\mathbf{r}$ and $\mathbf{v}$ are independent variables and eq.(9) is identically null for any arbitrary $\mathbf{v}$, the idempotent coefficients of $\mathbf{v}$ must be zero. In this way, we find the coefficient equation for the zero power of $\mathbf{v}$,

$$\frac{e\nabla\varphi\cdot\mathbf{u}}{kT}\sum_{i=0}^{\infty}\left[\frac{(1-q)m}{2kT}\right]^i u^{2i} = 0$$

and then we obtain

$$\nabla\varphi\cdot\mathbf{u} = 0. \quad (10)$$

In the following calculations we will use this result. The coefficient equation for the first power of $\mathbf{v}$ is

$$\nabla \ln\left[nB_q\left(\frac{m}{2\pi kT}\right)^{3/2}\right] - \frac{e}{kT}\left(\nabla\varphi - \frac{\mathbf{u}\times\mathbf{B}}{c}\right)$$
$$+\sum_{i=1}^{\infty}\left(\frac{m}{2kT}\right)^i u^{2i}\left[\mathbf{A}_{qi} - (1-q)^i\frac{e}{kT}\left(\nabla\varphi - \frac{\mathbf{u}\times\mathbf{B}}{c}\right)\right] = 0. \quad (11)$$

Generally, the coefficient equation of the $l$th power of $\mathbf{v}$, $l = 2, 3, 4, 5, ...,$ can be written as an unified form,

$$\mathbf{v} \cdot \sum_{i=0}^{\infty}\left(\frac{m}{2kT}\right)^i V_{l-1}(i, u, v^{l-1}) = 0, \quad (12)$$

where $V_l(i, v_0, v^l)$ denotes a sum of those terms containing $v^l$ in eq.(12), i.e.

$$V_l(i, u, v^l) = \sum_{\substack{p+s+j=i\\(s+2p=l)}}\frac{i!}{p!s!j!}(-2\mathbf{v}\cdot\mathbf{u})^s u^{2j}v^{2p}$$
$$\left[\mathbf{A}_{qi} - (1-q)^i\frac{e}{kT}\left(\nabla\varphi - \frac{\mathbf{u}\times\mathbf{B}}{c}\right)\right]. \quad (13)$$

First, from eq. (12) we can conclude a common equation satisfied by all the coefficient equations for $l = 2, 3, 4, 5, ...,$

$$\mathbf{A}_{qi} - (1-q)^i \frac{e}{kT}\left(\nabla\varphi - \frac{1}{c}\mathbf{u}\times\mathbf{B}\right) = 0 \cdot \quad (14)$$

And then, using eq.(8) we find the equation of the $q$-parameter:

$$k\nabla T - (1-q)e\left(\nabla\varphi - \frac{1}{c}\mathbf{u}\times\mathbf{B}\right) = 0 \cdot \quad (15)$$

From eq.(15) we can obtain an expression of the nonextensive parameter $q$ in the plasma with magnetic field. Eq.(15) tell us clearly that $q \ne 1$ holds in the plasma if and only if the temperature gradient is $\nabla T \ne 0$, and thus the power-law $q$-distribution (2) represents a stationary nonequilibrium distribution of the plasma. We therefore conclude that the power-law $q$-distribution (2) can describe the nonisothermal nature in the nonequilibrium plasmas with the electromagnetic interactions. The Maxwellian velocity distribution (for $q=1$) is a thermal equilibrium distribution and it is only a rough description for such plasmas. For the magnetic-field-free case, removing the magnetic field term in

eq.(15), it exactly reduces to the equation of the $q$-parameter given in the case of the plasmas without magnetic field [14].

Second, we may write a specific form of the generalized Maxwellian $q$-distribution of electrons in the plasma system. From eq.(11) we have

$$\nabla \ln\left[nB_q\left(\frac{m}{2\pi kT}\right)^{3/2}\right] - \frac{e}{kT}\left(\nabla\varphi - \frac{\boldsymbol{u}}{c}\times\boldsymbol{B}\right) = 0, \quad (16)$$

and then, we find the density distribution,

$$n(\boldsymbol{r}) = n_0\left(\frac{T(\boldsymbol{r})}{T_0}\right)^{3/2}\exp\left[\frac{e}{k}\left(\int_0^{\boldsymbol{r}}\frac{\nabla\varphi(\boldsymbol{r}')-(\boldsymbol{u}/c)\times\boldsymbol{B}(\boldsymbol{r}')}{T(\boldsymbol{r}')}\cdot d\boldsymbol{r}'\right)\right], \quad (17)$$

where the integral constants $n_0$ and $T_0$ denote the electron number density and the temperature at $\boldsymbol{r} = 0$, respectively. Substituting eq.(17) into eq.(2), we obtain the generalized Maxwellian $q$-distribution of the electrons as

$$f_q(\boldsymbol{r},\boldsymbol{v}) = n_0 B_q\left(\frac{m}{2\pi kT(\boldsymbol{r})}\right)^{3/2}\left[1-(1-q)\frac{m(\boldsymbol{v}-\boldsymbol{u})^2}{2kT(\boldsymbol{r})}\right]^{1/(1-q)}$$
$$\exp\left[\frac{e}{k}\left(\int_0^{\boldsymbol{r}}\frac{\nabla\varphi(\boldsymbol{r}')-(\boldsymbol{u}/c)\times\boldsymbol{B}(\boldsymbol{r}')}{T(\boldsymbol{r}')}\cdot d\boldsymbol{r}'\right)\right]. \quad (18)$$

If there is $\nabla T = 0$ in the plasma, we have $q = 1$, and then eq.(18) reduces to the Maxwellian velocity distribution in an electromagnetic field. As a power-law velocity distribution, this generalized Maxwellian $q$-distribution of the electrons in a magnetic field can be applied to the velocity distribution of the observed electrons within the plasma sheet in the magnetosphere [56], and the plasma velocity distributions with a "suprathermal" power-law tail at high energies discovered both in the solar wind and in planetary magnetospheres and magnetosheaths [57].

**Conclusion**. – In conclusion, we have studied the nonextensive parameter $q$ of the nonequilibrium electron gas in a magnetic field. We exactly obtained an equation of the $q$-parameter, given by eq. (15), based on the generalized Boltzmann equation, the $q$–H theorem and the power-law velocity $q$-distribution. It is shown that the $q$-parameter different from unity is closely related to the temperature gradient, the electric field strength, the magnetic induction as well as the overall bulk velocity of the plasma. The effect of the magnetic field on the $q$-parameter depends on the overall bulk velocity of the plasma. The whole analyses in the context followed the idea of [14], undoubtedly based on the standard line in the Boltzmann kinetic theory for plasmas, where the Coulombian interactions and Lorentz forces play dominant roles. The physical explanation of the parameter $q\neq 1$ for the plasma in magnetic field was presented clearly by eq.(15), which told us that it represents the nonequilibrium nature or the nonisothermal configurations of the nonequilibrium plasmas with the electromagnetic interactions. We also showed that the $q$-distributions such as (18) can be applied to study the characteristics of the astrophysical and space plasmas with the power-law kappa velocity distributions.

\*\*\*

This work is supported by the National Natural Science Foundation of China under grant No 11175128, and also by the Higher School Specialized Research Fund for Doctoral Program under grant No 20110032110058.